\documentclass[aps,prd,showpacs,11pt,tightenlines,preprint,superscriptaddress,nofootinbib]{revtex4-1}
\pdfoutput=1
\usepackage{amsfonts}
\usepackage{amsmath}
\usepackage{bbm}
\usepackage[T1]{fontenc}
\usepackage{graphicx}
\usepackage{hyperref}
\usepackage{multirow}
\usepackage[table]{xcolor}
\usepackage{xspace}
\usepackage{url}

\usepackage[normalem]{ulem}

\newcommand{\HEPfit}{\texttt{HEPfit}\xspace}

\definecolor{cambridgeblue}{rgb}{0.2, 0.2, 0.58}
\definecolor{darkraspberry}{rgb}{0.53, 0.15, 0.34}
\definecolor{darkyellow}{rgb}{0.9, 0.75, 0.0}

\newcommand*\cellr{\cellcolor{red!40!white}}
\newcommand*\celly{\cellcolor{darkyellow!45!white}}
\newcommand*\cellg{\cellcolor{green!45!white}}

\newcommand\redout{\bgroup\markoverwith
{\textcolor{red}{\rule[0.5ex]{2pt}{0.8pt}}}\ULon}

\newcommand\blueout{\bgroup\markoverwith
{\textcolor{blue}{\rule[0.5ex]{2pt}{0.8pt}}}\ULon}

\hypersetup{pdfstartview=FitV,colorlinks=true,linkcolor=darkraspberry,citecolor=cambridgeblue,filecolor=black,urlcolor=cambridgeblue}

\definecolor{color1}{HTML}{00dd00}
\definecolor{color2}{HTML}{dd0000}
\definecolor{color3}{HTML}{000000}
\definecolor{color4}{HTML}{000000}
\definecolor{color5}{HTML}{000000}
\definecolor{color6}{HTML}{000000}
\definecolor{color7}{HTML}{000000}
\definecolor{color8}{HTML}{000000}
\definecolor{color9}{HTML}{000000}
\definecolor{color10}{HTML}{000000}
\definecolor{color11}{HTML}{000000}
\definecolor{color12}{HTML}{000000}
\definecolor{color8TeV}{HTML}{ddddee}
\definecolor{color13TeV}{HTML}{eeee99}
\definecolor{colornew}{HTML}{FF99FF}

\begin{document}

\title{Global fits in the Georgi-Machacek model}

\author{Cheng-Wei Chiang}
\email{chengwei@phys.ntu.edu.tw}

\affiliation{Department of Physics, National Taiwan University, Taipei 10617, Taiwan}
\affiliation{Institute of Physics, Academia Sinica, Taipei 11529, Taiwan}

\author{Giovanna Cottin}
\email{gcottin@phys.ntu.edu.tw}

\affiliation{Department of Physics, National Taiwan University, Taipei 10617, Taiwan}

\author{Otto Eberhardt}
\email{otto.eberhardt@ific.uv.es}

\affiliation{IFIC, Universitat de Val\`{e}ncia - Consejo Superior de Investigaciones Cient\'ificas, Apt.~Correus 22085, E-46071, Val\`encia, Spain}

\date{\today}

\begin{abstract}
Off the beaten track of scalar singlet and doublet extensions of the Standard Model, triplets combine an interesting LHC phenomenology with an explanation for neutrino masses. The Georgi-Machacek model falls into this category, but it has never been fully explored in a global fit. We use the {\texttt{HEPfit}} package to combine recent experimental Higgs data with theoretical constraints and obtain strong limits on the mixing angles and mass differences between the heavy new scalars as well as their decay widths. We also find that the current signal strength measurements allow for a Higgs to vector boson coupling with an opposite sign to the Standard Model, but this possibility can be ruled out by the lack of direct evidence for heavy Higgs states. For these hypothetical particles, we identify the dominant decay channels and extract bounds on their branching ratios from the global fit, which can be used to single out the decay patterns relevant for the experimental searches.
\end{abstract}

\maketitle

\section{Introduction}
\label{sec:intro}

The discovery of a new scalar resonance at the LHC \cite{Aad:2012tfa,Chatrchyan:2012xdj}, 
consistent with the Higgs boson of the Standard Model (SM), confirms its particle content. Still
several experimental observations, such as data on neutrino oscillations \cite{Patrignani:2016xqp},
beg for new physics explanations, whose effects are actively being looked for by the LHC experiments.

Among the well-motivated directions for new physics beyond the SM is the presence of an extended Higgs sector, which can lead to richer Higgs phenomenology at colliders. One possibility is the existence of additional Higgs triplet representations of $SU(2)$, in which neutrino masses can arise from the interaction of the SM Higgs doublet with the triplet field, that acquires a vacuum expectation value (VEV) $v_{\Delta}$ after electroweak symmetry breakdown (EWSB) \cite{Schechter:1980gr,Cheng:1980qt}. Particularly in order to avoid conflicts with the electroweak $\rho$ parameter \cite{Patrignani:2016xqp}, the Georgi-Machacek (GM) model \cite{Georgi:1985nv,Chanowitz:1985ug} adds one complex and one real scalar triplet in a way that ensures custodial $SU(2)_{V}$ symmetry is preserved in the scalar potential after the EWSB. The model predicts the existence of several Higgs multiplets, whose mass eigenstates form a quintet ($H_{5}$), one triplet ($H_{3}$) and two singlets ($H_{1}$ and $h$) under the custodial symmetry. In this work, we denote the 125-GeV Higgs boson by $h$.

The rich Higgs particle spectrum and associated attractive phenomenology deserve in-depth studies, as it is of crucial importance to understand to which extent there is still room for new physics in the Higgs sector. Notably, if $v_{\Delta}$ is sufficiently large, we can have enhanced couplings between the SM-like Higgs boson and the weak gauge bosons. For example, $\kappa_W = 1.28^{+0.18}_{-017}$ is reported in a recent measurement by the CMS Collaboration~\cite{Sirunyan:2018egh}, giving a hint for us to consider a Higgs sector with larger field representations \cite{Gunion:1989ci,Chiang:2013rua}. The GM model serves as a minimal model with this feature. Modifications to the SM-like Higgs couplings with other particles can be probed by precise determination of the Higgs signal strengths at the LHC. Aside from loop-mediated processes, such data can constrain $v_{\Delta}$ and the mixing angle between the singlets $\alpha$ without the need to specify the heavy Higgs masses. In view of expected high precision in determining the Higgs couplings to other SM particles, Refs.~\cite{Chiang:2017vvo,Chiang:2018xpl} recently even computed the renormalized $\kappa$ factors, defined to be Higgs couplings in the model normalized to their corresponding SM values, at the one-loop level.  Since $H_1$ and $h$ are related via an orthogonal rotation, these signal strengths also provide significant constraints on the couplings of $H_{1}$ to SM particles.

Earlier studies had shown various collider constraints on the parameter space of the GM model \cite{Chiang:2012cn,Chiang:2013rua,Chiang:2014bia,Chiang:2015kka,Logan:2015xpa,Chiang:2015amq,Degrande:2017naf,Logan:2017jpr}. In Ref.~\cite{Chiang:2015amq}, for example, it was shown that after considering theoretical bounds (namely, the stability of the potential and perturbative unitarity at tree level), the LHC Higgs signal strengths, together with electroweak precision observables, a favored region in the $(v_{\Delta},\alpha)$ plane is chosen by the data.

In this work, we go beyond the existing literature by performing global parameter fits in the GM model, including up-to-date experimental results from Run 1 and Run 2 of the LHC, by making use of the \HEPfit open-source package \cite{HEPfit}. This approach is in stark contrast to studies that only examine specific benchmark scenarios (that may miss interesting possibilities), as all the model parameters are varied simultaneously in the fits and a model likelihood is obtained. The package also allows the possibility to identify which of the experimental data impose most stringent bounds.

This paper is organized as follows. We start with a brief review of the GM model in Sec.~\ref{sec:model}. A short overview over the \HEPfit package can be found in Sec.~\ref{sec:hepfit}. Theoretical constraints on the scalar potential stability and perturbative unitarity at tree level are included in our fits as described in Sec.~\ref{sec:theoryconstraints}. We then consider all available experimental data on Higgs boson signal strengths in Sec.~\ref{sec:signalstrengths}, including the $\gamma\gamma$ and the $Z\gamma$ modes, thus extending the considerations in Ref.~\cite{Chiang:2015amq}. Constraints from eighty heavy Higgs direct searches at the LHC are described in Sec.~\ref{sec:directsearches}. They are included in our Bayesian analysis, and greatly extend the amount of constraints analyzed in previous works \cite{Chiang:2015amq,Logan:2015xpa,Chiang:2014bia}. Combined results of the fits and discussions are presented in Sec.~\ref{sec:results}. We close the paper with a summary of our findings in Sec.~\ref{sec:summary}.

\section{The Georgi-Machacek model}
\label{sec:model}

In the GM model \cite{Georgi:1985nv,Chanowitz:1985ug}, $SU(2)$-triplet complex scalar $\chi$ and real scalar $\xi$ are added to the SM particle content. Assuming that the custodial symmetry is preserved at tree level, we can write the SM doublet and new triplet scalar fields as a bi-doublet and a bi-triplet, respectively,
\begin{align*}
\Phi &= \left( \begin{array}{cc} \left( \phi^0 \right)^* & \phi^+ \\ - \left( \phi^+ \right)^* & \phi^0 \end{array} \right), \qquad
\Delta = \left( \begin{array}{ccc} \left( \chi^0 \right)^* & \xi^+ & \chi^{++} \\ - \left( \chi^+ \right)^* & \xi^0 & \chi^+ \\ \left( \chi^{++} \right)^* & -\left( \xi^+ \right)^* & \chi^0 \end{array} \right).
\end{align*}
After EWSB, the scalar fields have the VEV's given by
\begin{align}
\langle\Phi\rangle
&= \frac{v_\Phi}{\sqrt2} {\mathbbm 1}_{2\times 2}
~~\mbox{and}~~
\langle\Delta\rangle
= v_\Delta {\mathbbm 1}_{3\times 3}~.
\end{align}

Using, the above-defined fields, the scalar potential reads
\begin{align}
V(\Phi,\Delta) = & \frac12 m_{\Phi}^2 {\rm tr} \left[ \Phi^\dagger \Phi \right] + \frac12 m_{\Delta}^2 {\rm tr} \left[ \Delta^\dagger \Delta \right] + \lambda_1 \left( {\rm tr} \left[ \Phi^\dagger \Phi \right] \right)^2 + \lambda_2 \left( {\rm tr} \left[ \Delta^\dagger \Delta \right] \right)^2 \nonumber \\
& + \lambda_3 {\rm tr} \left[ \left( \Delta^\dagger \Delta \right)^2 \right] + \lambda_4 {\rm tr} \left[ \Phi^\dagger \Phi \right] {\rm tr} \left[ \Delta^\dagger \Delta \right] + \lambda_5 {\rm tr} \left[ \Phi^\dagger \frac{\sigma^a}{2} \Phi \frac{\sigma^b}{2} \right] {\rm tr} \left[ \Delta^\dagger T^a \Delta T^b \right] \nonumber \\
& + \mu_1 {\rm tr} \left[ \Phi^\dagger \frac{\sigma^a}{2} \Phi \frac{\sigma^b}{2} \right] (P^\dagger \Delta P)_{ab} + \mu_2 {\rm tr} \left[ \Delta^\dagger T^a \Delta T^b \right] (P^\dagger \Delta P)_{ab} 
~,
\label{eq:GMpot}
\end{align}
where $\sigma^a$ are the Pauli matrices, $T^a$ are the $3\times3$ matrix representation of the $SU(2)$ generators, and the similarity transformation relating the $SU(2)$ generators in the triplet and adjoint representations is given by
\begin{align*}
P = & \frac{1}{\sqrt{2}} \left( \begin{array}{ccc} -1 & i & 0 \\ 0 & 0 & \sqrt{2} \\ 1 & i & 0 \end{array} \right).
\end{align*}
Note that the triplet VEV is induced by the SM EWSB via the $\mu_1$ interaction.

Under the custodial $SU(2)_V$ symmetry, the physical eigenstates can be written as a quintet $H_5 = (H_5^{++}, H_5^+, H_5^0, H_5^-, H_5^{--})^T$ with mass $m_5$, a triplet $H_3 = (H_3^+, H_3^0, H_3^-)^T$ with mass $m_3$ and two scalar singlets $H_1$ and $h$, of which the former has the mass $m_{1}$ and the latter is identified with the $125$ GeV scalar boson found at the LHC. The relations between the physical fields and the original fields can be found in, for example, Ref.~\cite{Chiang:2012cn}. Rotating from the original basis to the mass basis involves two mixing angles $\alpha$ and $\beta$, where $\alpha$ diagonalizes the singlet subspace and $\tan \beta \equiv v_\Phi/(2\sqrt{2}v_\Delta)$ is used in the diagonalization of the Goldstone modes and the physical triplet states. In the limit of custodial symmetry, the states in each of the above-mentioned representations are degenerate in mass. An ${\cal O}(100)$~MeV mass splitting is expected among the states within the same representation because of custodial symmetry breaking by hypercharge interactions. In this article, we assume that $h$ is the lightest scalar boson in the GM Higgs spectrum.

We list a few remarkable features of the GM model here.  First, the $hWW$ and $hZZ$ couplings can be larger than the SM values at tree level. This does not happen in models extended with only singlet and/or doublet scalars. This feature is resistant to loop corrections, as explicitly shown in Refs.~\cite{Chiang:2017vvo,Chiang:2018xpl} at the one-loop level. Secondly, the quintet Higgs bosons have couplings with the weak gauge bosons, while the triplet Higgs bosons do not. The triplet Higgs bosons are thus said to be gauge-phobic. On the other hand, the triplet Higgs bosons have couplings with SM fermions, while the quintet Higgs bosons do not. The latter are thus said to be fermiophobic. Finally, the $H_5^0ZZ$ coupling divided by the $H_5^0WW$ coupling is $-2$, while the corresponding ratios for $h$ and $H_1$ are $1$.

\section{HEPfit}
\label{sec:hepfit}

The open-source package \HEPfit is a multi-purpose tool to calculate many different high-energy physics observables and theory constraints in various models. It is interfaced 
with {\texttt {BAT}}~\cite{Caldwell:2008fw} to perform Bayesian fits with Markov Chain Monte 
Carlo simulations. Here, we present the first results from the implementation of the GM model into \HEPfit. The global fit allows us to scrutinize this model with unprecedented precision as it allows us to vary all GM parameters simultaneously, and thus guarantees that we do not miss important features when scanning over the parameter space. This method has also been used in the two-Higgs doublet model~\cite{Cacchio:2016qyh,Chowdhury:2017aav,Eberhardt:2018lub}, and the GM implementation is partially based on the well-tested two-Higgs doublet model part of \HEPfit in order to minimize possible sources of errors.
We also cross-checked some benchmark points with the public code \texttt{GMcalc} \cite{Hartling:2014xma}. At tree-level we found agreement on all couplings. Concerning the scalar couplings to $\gamma\gamma$ and $Z\gamma$ we observe deviations due to the different implementation of higher-order corrections. Moreover, \HEPfit does not contain the one-loop decays $H_{3,5}^+\to W^+ \gamma$.

In our fits, we fix $m_h=125.09$ GeV \cite{Khachatryan:2016vau} and $v=\sqrt{v_\Phi^2+8v_\Delta^2}\approx 246$~GeV and all other SM parameters to their best-fit values~\cite{deBlas:2016ojx}. We use the following prior ranges for the remaining GM parameters:
\begin{align*}
150\text{ GeV } & \leq m_{1},m_{3},m_{5} \leq 1100\text{ GeV},\\
0\text{ GeV } & \leq v_\Delta \leq 86\text{ GeV},\\
-90^\circ & \leq \alpha \leq 90^\circ,\\
-1500\text{ GeV } & \leq \mu_1,\mu_2 \leq 1500\text{ GeV},\\
\end{align*}
where the masses $m_{1},m_{3},m_{5}$ of the $H_{1}$, $H_{3}$ and $H_{5}$ bosons, respectively, are chosen to be heavier than the $125$~GeV Higgs and lighter than 1.1 TeV, as we want to cover the ranges that are interesting for the LHC searches of heavy scalars. Accordingly, we also limit the absolute values of the trilinear couplings $\mu_1$ and $\mu_2$ to be below 1.5 TeV.

Concerning the heavy masses $m_{1}$, $m_{3}$ and $m_{5}$, our type of priors will depend on the set of constraints being used. For the direct searches, we will use flat mass priors, as the search limits depend on the masses linearly. As for the $h$ signal strengths and the theory bounds, they depend on the squared masses.  Therefore, we choose flat priors for $m_{1}^2$, $m_{3}^2$ and $m_{5}^2$ between $(150$ GeV$)^2$ and $(1100$ GeV$)^2$ in this case. In the global fit to all constraints, we apply both types of priors in two separate fits and overlay both fits in the figures and for the extraction of the limits. (See also Appendix~B of Ref.~\cite{Chowdhury:2017aav} for the same procedure in two-Higgs doublet model fits.)

\section{Fit constraints}
\label{sec:constraints}

In this section, we list the theoretical and experimental constraints imposed on 
the GM model parameter space in this analysis.

\subsection{Theory constraints}
\label{sec:theoryconstraints}

We take into account two different sets of theoretical constraints: stability of the scalar potential 
and perturbative unitarity, both at tree level. Stability of the electroweak vacuum is imposed by requiring that the scalar potential be bounded from below, which places restrictions on the $\lambda$ quartic couplings. We implement the constraints from Section~4 of Ref.~\cite{Arhrib:2011uy}.


Perturbative unitarity of the $S$-matrix of 2 scalars to 2 scalars scattering processes forces additional restrictions on the quartic couplings. We implement all seventeen constraints from the full $S$-matrix described in Ref.~\cite{Aoki:2007ah}.  Here we take the stronger limits that the real parts of the zeroth partial wave amplitudes have absolute values of less than $1/2$.


We note that the theoretical bounds implemented in this work are conservative. Perturbative unitarity can be broken in the GM model~\cite{Krauss:2018orw}. Also, the tree-level vacuum stability constraints can change once loop corrections are included~\cite{Krauss:2017xpj}. Since the focus of this work is on LHC constraints, we keep a more relaxed analysis in terms of the allowed parameter space from the theory side, but this can be more restrictive.

While the theory constraints are defined in terms of the quartic couplings of the scalar potential in Eq.~\eqref{eq:GMpot}, the following experimental bounds constrain the physical masses and the couplings of the scalars.

\subsection{Higgs signal strengths}
\label{sec:signalstrengths}

\begin{table}
\begin{center}
\begin{tabular}{lr|c|c|c|c|c|c|c|}
& & $b\bar b$  & $WW$ & $\tau\tau$ & $ZZ$ & $\gamma\gamma$ & $Z\gamma$ & $\mu\mu$ \\
\hline
\hline
\multicolumn{2}{r|}{SM Br} & 57.5\% & 21.6\% & 6.3\% & 2.7\% & 2.3\textperthousand & 1.6\textperthousand & 0.2\textperthousand \\
\hline
\hline
ggF$_8$ &87.2\%&--&\cellg \cite{ATLAS:2014aga,Chatrchyan:2013iaa}&\celly \cite{Aad:2015vsa,Chatrchyan:2014nva}&\cellg \cite{Aad:2014eva,Khachatryan:2014jba}&\cellg \cite{Aad:2014eha,Khachatryan:2014ira}&\cellr \cite{Aad:2015gba,Chatrchyan:2013vaa} &\cellr \cite{Khachatryan:2016vau}\\
ggF$_{13}$ &87.1\%&--&\cellg \cite{ATLAS-CONF-2018-004,Sirunyan:2018egh}&\celly \cite{ATLAS-CONF-2018-021,Sirunyan:2017khh}&\cellg \cite{ATLAS-CONF-2018-018,Sirunyan:2017exp,CMS-PAS-HIG-18-001}&\cellg \cite{ATLAS-CONF-2018-028,Sirunyan:2018ouh}&\cellr \cite{Aaboud:2017uhw,Sirunyan:2018tbk} &\cellr \cite{ATLAS-CONF-2018-026,Sirunyan:2018hbu}\\
\cline{1-7}
VBF$_8$ &7.2\%&--&\cellg \cite{ATLAS:2014aga,Chatrchyan:2013iaa}&\celly \cite{Aad:2015vsa,Chatrchyan:2014nva}&\cellr \cite{Aad:2014eva,Khachatryan:2014jba}&\celly \cite{Aad:2014eha,Khachatryan:2014ira}&\cellr &\cellr \\
VBF$_{13}$ &7.4\%&\cellr \cite{Aaboud:2018gay,CMS-PAS-HIG-16-003}&\cellg \cite{ATLAS-CONF-2018-004,Sirunyan:2018egh}&\celly \cite{ATLAS-CONF-2018-021,Sirunyan:2017khh}&\celly \cite{ATLAS-CONF-2018-018,Sirunyan:2017exp,CMS-PAS-HIG-18-001}&\cellg \cite{ATLAS-CONF-2018-028,Sirunyan:2018ouh}&\cellr &\cellr \\
\cline{1-7}
Vh$_8$ &5.1\%&\celly \cite{Aad:2014xzb,Chatrchyan:2013zna}&\cellr \cite{Aad:2015ona,Chatrchyan:2013iaa}&\cellr \cite{Aad:2015vsa,Chatrchyan:2014nva}&\cellr \cite{Aad:2014eva,Khachatryan:2014jba}&\celly \cite{Aad:2014eha,Khachatryan:2014ira}&\cellr &\cellr \\
Vh$_{13}$ &4.4\%&\cellg \cite{ATLAS-CONF-2018-036,Sirunyan:2017elk}&\cellr \cite{ATLAS-CONF-2016-112,Sirunyan:2018egh}&\cellr \cite{Sirunyan:2017khh,Sirunyan2018aaa}&\celly \cite{ATLAS-CONF-2018-018,Sirunyan:2017exp,CMS-PAS-HIG-18-001}&\celly \cite{ATLAS-CONF-2018-028,Sirunyan:2018ouh}&\cellr &\cellr\\
\cline{1-7}
tth$_8$ &0.6\%&\cellr \cite{Aad:2015gra,Khachatryan:2014qaa}&--&--&\cellr \cite{Aad:2014eva,Khachatryan:2014jba}&\cellr \cite{Aad:2014eha,Khachatryan:2014ira}&\cellr &\cellr \\
tth$_{13}$ &1.0\%&\cellg \cite{Aaboud:2017rss,Sirunyan:2018ygk,Sirunyan:2018hoz}&\celly \cite{Aaboud:2017jvq,Sirunyan:2018egh,Sirunyan:2018shy}&\cellr \cite{Aaboud:2017jvq,Sirunyan:2018shy}&\cellr \cite{ATLAS-CONF-2018-018,Aaboud:2017jvq,Sirunyan:2017exp,Sirunyan:2018shy,CMS-PAS-HIG-18-001}&\cellg \cite{ATLAS-CONF-2018-028,Sirunyan:2018ouh}&\cellr &\cellr \\
\hline
\hline
Vh$_2$ & & \celly \cite{Aaltonen:2013ipa,Abazov:2013gmz} \\
\hline
tth$_2$ & & \cellr \cite{Aaltonen:2013ipa} \\
\hline
\end{tabular}\\[10pt]
\begin{tabular}{|c||c||c|}\hline\cellg $0<\hat\sigma<0.5$ &\celly $0.5\leq \hat\sigma \leq 1.0$ &\cellr $\hat\sigma>1.0$\\\hline\end{tabular} \quad ($\hat\sigma=\sigma_{\text{\tiny{min}}}/w$)
\caption{Higgs signal strength inputs used in our fits. The Higgs decays are listed in separate columns, with the corresponding SM branching ratios given in the second line. In lines three to twelve, we give all LHC and Tevatron references of the used signal strengths, ordered by production mechanism and $\sqrt{s}$. For the LHC, we indicate the share of Higgs production in $pp$ collisions for each channel in the second column. The background colors of the table cells give an idea about how precise the strongest signal strength measurement for a particular production mechanism is at present: green cells contain results with an uncertainty of less than 0.5 on $\mu$, yellow cells have an uncertainty between 0.5 and 1, and red entries have not been measured with a precision smaller than 1 (see the text for more details). On the decays to $Z\gamma$ and $\mu\mu$, we only have information for $pp$ production and assume the SM composition in the second column for them.}
\label{tab:signalstrengthinputs}
\end{center}
\end{table}

For the signal strengths computation, the predicted SM Higgs production cross-section $\sigma$ and total decay width $\Gamma$ are dressed with scale factors. For the production modes $i=$ ggF, VBF, Wh, Zh, tth and the decay modes $f=ZZ, WW, \gamma\gamma, Z\gamma, \tau\tau, \mu\mu, b\bar{b}$, we define $r_{i}$ and $r_{f}$ to be respectively the ratios of the production cross section $\sigma_{i}$ and the decay width $\Gamma_{f}$ with respect to their corresponding SM values.  Therefore, the production cross section times the branching ratio for a particular channel in the GM model is given by
\begin{equation}
\left(\sigma_{i} \cdot \mathcal{B}_{f}\right)_{\mbox{\tiny{GM}}} = \left(\sigma_{i} \cdot \mathcal{B}_{f} \right)_{\mbox{\tiny{SM}}} \cdot r_{i}\cdot r_{f} \cdot \frac{\Gamma_{\mbox{\tiny{SM}}}}{\Gamma_{\mbox{\tiny{GM}}}},
\end{equation}
with $\Gamma_{\mbox{\tiny{SM}}}$ and $\Gamma_{\mbox{\tiny{GM}}}$ being the total widths of the Higgs boson in the SM and the GM model, respectively.

To quantify the deviation of the GM model from the SM, the signal strength of a process $\mu_{i}^{f}$ 
with the production channel $i$ and the decay of $h$ to an $f$ final state is then defined as
\begin{equation}
\mu_{i}^{f}=\frac{r_{i}\cdot r_{f}}{\sum^{}_{f'} r_{f'}\cdot \mathcal{B}_{\mbox{\tiny {SM}}}(h\rightarrow f')}.
\end{equation}
Each signal strength is computed in the narrow-width approximation, and depends on the GM $h$ couplings to all final states. The values for all couplings are cross-checked with the predictions in Ref.~\cite{Hartling:2014xma}. 

The experimental input values of the Higgs signal strengths are similar to the ones in Ref.~\cite{deBlas:2018tjm}, only that we updated some numbers after the ICHEP 2018. Instead of all 138 numerical signal strength inputs, we show in Table~\ref{tab:signalstrengthinputs} the current sensitivity of the individual channels, indicated by the background colors. The quantity $\hat\sigma$ is the ratio of the smallest uncertainty of all individual measurements in one table cell ($\sigma_{\text{\tiny{min}}}$) and the weight of the corresponding production mechanism ($w$). For instance, in Ref.~\cite{Sirunyan:2017khh}, we can find that $\mu^{\tau\tau}=1.11^{+0.34}_{-0.35}$ in their ``VBF'' category, so $\sigma_{\text{\tiny{min}}}=0.34$ here. Note that the categories do not consist of only one production mechanism, and thus the given value is no measurement of $\mu^{\tau\tau}_\text{VBF}$. The admixture (weight) of VBF is only 57\%, and so $\hat\sigma\approx 0.6$ in this case. We stress that $\hat\sigma$ depends on the \textit{individual} measurements and not on the combination. It is only intended to give the reader a rough estimate of the achieved precision in every channel, and should not be understood as a quantitative statement. In the last two columns, we use the 8-TeV data from Refs.~\cite{Aad:2015gba,Chatrchyan:2013vaa} (\cite{Khachatryan:2016vau}) and the 13-TeV results from Refs.~\cite{Aaboud:2017uhw,Sirunyan:2018tbk} (\cite{ATLAS-CONF-2018-026,Sirunyan:2018hbu}) for the $Z\gamma$ ($\mu\mu$) final state, since the only information about the initial state is the inclusive $pp$ production rather than individual channels.

\subsection{Searches for heavy Higgs particles}
\label{sec:directsearches}

\begin{table}
\begin{center}
\begin{tabular}{| l | l | l  l | c | c | c |}
\hline
\textbf{Label} &\textbf{Channel} & \multicolumn{2}{| l |}{\textbf{Experiment}} & \textbf{Mass range} & ${\cal L}$ \\
               &                 & \multicolumn{2}{| l |}{} &  \textbf{[TeV]} & \textbf{[fb$^{-1}$]} \\[1pt]
\hline
\hline
\cellcolor{color13TeV} $A_{13t}^{tt}$  & $tt \to \phi^0 \to tt$ & ATLAS & \cite{Aaboud:2018xpj} & [0.4;1] & 36.1 \\
\hline
\cellcolor{color13TeV} $A_{13b}^{tt}$  & $bb \to \phi^0 \to tt$ & ATLAS & \cite{ATLAS-CONF-2016-104} & [0.4;1] & 13.2\\
\hline
\hline
\cellcolor{color8TeV} $C_{8b}^{bb}$ &$bb \to \phi^0 \to bb$ & CMS & \cite{Khachatryan:2015tra} & [0.1;0.9] & 19.7\\
\hline
\cellcolor{color8TeV} $C_{8}^{bb}$  &$gg \to \phi^0\to bb$ & CMS & \cite{Sirunyan:2018pas} & [0.33;1.2] & 19.7\\
\hline
\cellcolor{color13TeV} $C_{13}^{bb}$ & $pp \to \phi^0\to bb$ & CMS & \cite{CMS-PAS-HIG-16-025} & [0.55;1.2] & 2.69\\
\hline
\cellcolor{color13TeV} $C_{13b}^{bb}$ & $bb \to \phi^0\to bb$ & CMS & \cite{Sirunyan:2018taj} & [0.3;1.3] & 35.7\\
\hline
\hline
\cellcolor{color8TeV} $A_{8}^{\tau\tau}$ &\multirow{2}{*}{$gg\to \phi^0 \to \tau\tau$} & ATLAS &\cite{Aad:2014vgg} & [0.09;1] & 20 \\
\cellcolor{color8TeV} $C_{8}^{\tau\tau}$  & & CMS &\cite{CMS-PAS-HIG-14-029} &  [0.09;1]  &19.7 \\
\hline
\cellcolor{color8TeV} $A_{8b}^{\tau\tau}$  &\multirow{2}{*}{$bb\to \phi^0 \to \tau\tau$} & ATLAS &\cite{Aad:2014vgg} & [0.09;1] & 20 \\
\cellcolor{color8TeV} $C_{8b}^{\tau\tau}$ & & CMS & \cite{CMS-PAS-HIG-14-029}& [0.09;1] & 19.7 \\
\hline
\cellcolor{color13TeV} $A_{13}^{\tau\tau}$ &\multirow{2}{*}{$gg \to \phi^0\to \tau \tau$} & ATLAS & \cite{Aaboud:2017sjh} & [0.2;2.25] & 36.1\\[-1pt]
\cellcolor{color13TeV} $C_{13}^{\tau\tau}$ &  & CMS & \cite{Sirunyan:2018zut} & [0.09;3.2] & 35.9\\
\hline
\cellcolor{color13TeV} $A_{13b}^{\tau\tau}$ &\multirow{2}{*}{$bb \to \phi^0\to \tau \tau$} & ATLAS & \cite{Aaboud:2017sjh} & [0.2;2.25] & 36.1\\[-1pt]
\cellcolor{color13TeV} $C_{13b}^{\tau\tau}$ & & CMS & \cite{Sirunyan:2018zut} & [0.09;3.2] & 35.9\\
\hline
\end{tabular}
\caption{Neutral heavy Higgs boson searches relevant for the GM scalars with fermionic final states. $\phi^0=H_1^0,H_3^0$.}
\label{tab:directsearchesA}
\end{center}
\end{table}

\begin{table}
\begin{center}
\begin{tabular}{| l | l | l  l | c | c | c |}
\hline
\textbf{Label} &\textbf{Channel} & \multicolumn{2}{| l |}{\textbf{Experiment}} & \textbf{Mass range} & ${\cal L}$ \\
               &                 & \multicolumn{2}{| l |}{} &  \textbf{[TeV]} & \textbf{[fb$^{-1}$]} \\[1pt]
\hline
\cellcolor{color8TeV} $A_{8}^{\gamma\gamma}$&$gg\to \phi^0 \to \gamma\gamma$ & ATLAS &\cite{Aad:2014ioa} & [0.065;0.6] & 20.3 \\
\hline
\cellcolor{color13TeV} $A_{13}^{\gamma\gamma}$ &$pp \to \phi^0\to \gamma \gamma$ & ATLAS & \cite{Aaboud:2017yyg} & [0.2;2.7] & 36.7\\
\hline
\cellcolor{color13TeV} $C_{13}^{\gamma\gamma}$ & $gg \to \phi^0\to \gamma \gamma$ & CMS & \cite{Khachatryan:2016yec} & [0.5;4] & 35.9\\ 
\hline
\hline
\cellcolor{color8TeV} $A_{8}^{Z\gamma}$  &\multirow{2}{*}{$pp\to \phi^0 \to Z\gamma \to (\ell \ell) \gamma$} & ATLAS & \cite{Aad:2014fha} & [0.2;1.6] & 20.3 \\
\cellcolor{color8TeV} $C_{8}^{Z\gamma}$ & & CMS & \cite{CMS-PAS-HIG-16-014} & [0.2;1.2] & 19.7 \\
\hline
\cellcolor{color13TeV} $A_{13}^{\ell\ell\gamma}$  & $gg \to \phi^0\to Z \gamma [\to (\ell \ell) \gamma ]$ & ATLAS & \cite{Aaboud:2017uhw} & [0.25;2.4] & 36.1\\
\hline
\cellcolor{color13TeV} $A_{13}^{qq\gamma}$  & $gg \to \phi^0\to Z \gamma [\to (qq) \gamma ]$ & ATLAS & \cite{Aaboud:2018fgi} & [1;6.8] & 36.1\\
\hline
\cellcolor{color13TeV} $C_{8+13}^{Z\gamma}$  & $gg \to \phi^0\to Z \gamma$ & CMS & \cite{Sirunyan:2017hsb} & [0.35;4] & 35.9\\
\hline
\hline
\cellcolor{color8TeV} $A_{8}^{ZZ}$  &$gg\to \phi^0\to ZZ$ & ATLAS & \cite{Aad:2015kna}& [0.14;1] & 20.3 \\
\hline
\cellcolor{color8TeV} $A_{8V}^{ZZ}$  &$VV \to \phi^0\to ZZ$ & ATLAS & \cite{Aad:2015kna}& [0.14;1] & 20.3 \\
\hline
\cellcolor{color13TeV} $A_{13}^{2\ell2L}$  & $gg\to \phi^0 \to ZZ [\to (\ell \ell) (\ell \ell, \nu \nu)]$ & ATLAS & \cite{Aaboud:2017rel} & [0.2;1.2] & 36.1\\
\hline
\cellcolor{color13TeV} $A_{13V}^{2\ell2L}$  & $VV\to \phi^0\to ZZ [\to (\ell \ell) (\ell \ell, \nu \nu)]$ & ATLAS & \cite{Aaboud:2017rel} & [0.2;1.2] & 36.1\\
\hline
\cellcolor{color13TeV} $A_{13}^{2L2q}$ & $gg\to \phi^0\to ZZ [\to (\ell \ell, \nu \nu) (qq)]$ & ATLAS & \cite{Aaboud:2017itg} & [0.3;3] & 36.1\\
\hline
\cellcolor{color13TeV} $A_{13V}^{2L2q}$ & $VV\to \phi^0\to ZZ [\to (\ell \ell, \nu \nu) (qq)]$ & ATLAS & \cite{Aaboud:2017itg} & [0.3;3] & 36.1\\
\hline
\cellcolor{color13TeV} $C_{13}^{2\ell2X}$ & $pp\to \phi^0\to ZZ [\to (\ell \ell) (qq,\nu\nu,\ell\ell)]$ & CMS & \cite{Sirunyan:2018qlb} & [0.13;3] & 35.9\\
\hline
\cellcolor{color13TeV} $C_{13}^{2q2\nu}$ & $pp\to \phi^0\to ZZ [\to (qq)(\nu\nu)]$ & CMS & \cite{Sirunyan:2018ivv} & [1;4] & 35.9\\
\hline
\hline
\cellcolor{color8TeV} $A_{8}^{WW}$  &$gg\to \phi^0\to WW$ & ATLAS &\cite{Aad:2015agg}& [0.3;1.5] & 20.3 \\
\hline
\cellcolor{color8TeV} $A_{8V}^{WW}$ &$VV \to \phi^0\to WW$ & ATLAS & \cite{Aad:2015agg}& [0.3;1.5] & 20.3 \\
\hline
\cellcolor{color13TeV} $A_{13}^{2(\ell\nu)}$  & $gg\to \phi^0\to WW [\to (e \nu) (\mu \nu)]$ & ATLAS & \cite{Aaboud:2017gsl} & [0.25;4] & 36.1\\
\hline
\cellcolor{color13TeV} $A_{13V}^{2(\ell\nu)}$  & $VV\to \phi^0\to WW [\to (e \nu) (\mu \nu)]$ & ATLAS & \cite{Aaboud:2017gsl} & [0.25;3] & 36.1\\
\hline
\cellcolor{color13TeV} $C_{13}^{2(\ell\nu)}$ & $(gg\!+\!VV)\to \phi^0\to WW \to (\ell \nu) (\ell \nu)$ & CMS & \cite{CMS-PAS-HIG-16-023} & [0.2;1] & 2.3\\
\hline
\cellcolor{color13TeV} $A_{13}^{\ell\nu2q}$ & $gg\to \phi^0\to WW[\to (\ell \nu) (qq)]$ & ATLAS & \cite{Aaboud:2017fgj} & [0.3;3] & 36.1\\
\hline
\cellcolor{color13TeV} $A_{13V}^{\ell\nu2q}$ & $VV\to \phi^0\to WW[\to (\ell \nu) (qq)]$ & ATLAS & \cite{Aaboud:2017fgj} & [0.3;3] & 36.1\\
\hline
\cellcolor{color13TeV} $C_{13}^{\ell\nu2q}$ & $pp\to \phi^0\to WW[\to (\ell \nu) (qq)]$ & CMS & \cite{Sirunyan:2018iff} & [1;4.4] & 35.9\\
\hline
\hline
\cellcolor{color8TeV} $C_{8}^{VV}$  & $pp \to \phi^0\to VV$ & CMS & \cite{Khachatryan:2015cwa} & [0.145;1] & 24.8 \\
\hline
\end{tabular}
\caption{Neutral heavy Higgs boson searches relevant for the GM scalars with vector boson final states. $\phi^0=H_1^0,H_3^0,H_5^0$ and $\ell=e,\mu$.}
\label{tab:directsearchesB}
\end{center}
\end{table}

\begin{table}
\begin{center}
\begin{tabular}{| l | l | l  l | c | c | c |}
\hline
\textbf{Label} &\textbf{Channel} & \multicolumn{2}{| l |}{\textbf{Experiment}} & \textbf{Mass range} & ${\cal L}$ \\
               &                 & \multicolumn{2}{| l |}{} &  \textbf{[TeV]} & \textbf{[fb$^{-1}$]} \\[1pt]
\hline
\hline
\cellcolor{color8TeV} $A_{8}^{hh}$ &$gg\to H^{0}_{1}\to hh$ & ATLAS &\cite{Aad:2015xja} & [0.26;1] & 20.3\\
\hline
\cellcolor{color8TeV} $C_{8}^{4b}$  &$pp\to H^{0}_{1}\to hh \to (bb) (bb)$ & CMS &\cite{Khachatryan:2015yea} & [0.27;1.1] & 17.9\\
\hline
\cellcolor{color8TeV} $C_{8}^{2\gamma2b}$  &$pp\to H^{0}_{1}\to hh \to (bb) (\gamma \gamma)$ & CMS & \cite{Khachatryan:2016sey} & [0.260;1.1] & 19.7\\
\hline
\cellcolor{color8TeV} $C_{8g}^{2b2\tau}$ &$gg\to H^{0}_{1}\to hh \to (bb) (\tau\tau)$ & CMS & \cite{Khachatryan:2015tha} & [0.26;0.35] & 19.7\\
\hline
\cellcolor{color8TeV} $C_{8}^{2b2\tau}$  &$pp\to H^{0}_{1}\to hh [\to (bb) (\tau\tau)]$ & CMS & \cite{Sirunyan:2017tqo} & [0.35;1] & 18.3\\
\hline
\cellcolor{color13TeV} $A_{13}^{4b}$& \multirow{2}{*}{$pp \to H^{0}_{1}\to hh \to (bb) (bb)$} & ATLAS & \cite{Aaboud:2018knk} & [0.26;3] & 36.1\\[-1pt]
\cellcolor{color13TeV} $C_{13}^{4b}$  & & CMS & \cite{Sirunyan:2018zkk} & [0.26;1.2] & 35.9\\
\hline
\cellcolor{color13TeV} $A_{13}^{2\gamma2b}$  & $pp \to H^{0}_{1}\to hh [\to (bb) (\gamma \gamma)]$ & ATLAS & \cite{Aaboud:2018ftw} & [0.26;1] & 36.1\\
\cellcolor{color13TeV} $C_{13}^{2\gamma2b}$  & $pp \to H^{0}_{1}\to hh \to (bb) (\gamma \gamma)$ & CMS & \cite{Sirunyan:2018iwt} & [0.25;0.9] & 35.9\\
\hline
\cellcolor{color13TeV} $A_{13}^{2b2\tau}$ & \multirow{2}{*}{$pp \to H^{0}_{1}\to hh \to (bb) (\tau \tau)$} & ATLAS & \cite{Aaboud:2018sfw} & [0.26;1] & 36.1\\
\cellcolor{color13TeV} $C_{13,1}^{2b2\tau}$ & & CMS & \cite{Sirunyan:2017djm} & [0.25;0.9] & 35.9\\
\cellcolor{color13TeV} $C_{13,2}^{2b2\tau}$ & $pp \to H^{0}_{1}\to hh [\to (bb) (\tau \tau)]$ & CMS & \cite{Sirunyan:2018fuh} & [0.9;4] & 35.9\\
\hline
\cellcolor{color13TeV} $C_{13}^{2b2V}$ & $pp \to H^{0}_{1}\to hh \to (bb) (VV\to \ell \nu \ell \nu)$ & CMS & \cite{Sirunyan:2017guj} & [0.26;0.9] & 35.9\\
\hline
\cellcolor{color13TeV} $A_{13}^{2\gamma2W}$  & $gg \to H^{0}_{1}\to hh \to (\gamma \gamma) (WW)$ & ATLAS & \cite{Aaboud:2018ewm} & [0.26;0.5] & 36.1\\ 
\hline
\hline
\cellcolor{color8TeV} $A_{8}^{bbZ}$ &$gg\to H^{0}_{3}\to hZ \to (bb) Z$ & ATLAS & \cite{Aad:2015wra} & [0.22;1] & 20.3\\
\hline
\cellcolor{color8TeV} $C_{8}^{2b2\ell}$ &$gg\to H^{0}_{3}\to hZ \to (bb) (\ell \ell)$ & CMS &\cite{Khachatryan:2015lba} & [0.225;0.6] &19.7\\
\hline
\cellcolor{color8TeV} $A_{8}^{\tau\tau Z}$ &$gg\to H^{0}_{3}\to hZ \to (\tau\tau) Z$ & ATLAS & \cite{Aad:2015wra} & [0.22;1] & 20.3\\
\hline
\cellcolor{color8TeV} $C_{8}^{2\tau2\ell}$  &$gg\to H^{0}_{3}\to hZ \to (\tau\tau) (\ell \ell)$ & CMS & \cite{Khachatryan:2015tha} & [0.22;0.35] & 19.7\\
\hline
\cellcolor{color13TeV} $A_{13}^{bbZ}$  & \multirow{3}{*}{$gg\to H^{0}_{3}\to hZ \to (bb) Z$} & ATLAS & \cite{Aaboud:2017cxo} & [0.2;2] & 36.1\\
\cellcolor{color13TeV} $C_{13,1}^{bbZ}$  & & CMS & \cite{CMS-PAS-HIG-18-005} & [0.22;0.8] & 35.9\\
\cellcolor{color13TeV} $C_{13,2}^{bbZ}$  & & CMS & \cite{Sirunyan:2018qob} & [0.8;2] & 35.9\\
\hline
\cellcolor{color13TeV} $A_{13b}^{bbZ}$  & \multirow{3}{*}{$bb\to H^{0}_{3}\to hZ \to (bb) Z$} & ATLAS & \cite{Aaboud:2017cxo} & [0.2;2] & 36.1\\
\cellcolor{color13TeV} $C_{13b,1}^{bbZ}$  & & CMS & \cite{CMS-PAS-HIG-18-005} & [0.22;0.8] & 35.9\\
\cellcolor{color13TeV} $C_{13b,2}^{bbZ}$  & & CMS & \cite{Sirunyan:2018qob} & [0.8;2] & 35.9\\
\hline
\hline
\cellcolor{color8TeV} $C_{8}^{\phi Z}$& $pp\to \phi^{0}\to \phi^{0\prime} Z \to (bb) (\ell\ell)$ & CMS & \cite{Khachatryan:2016are} & [0.13;1] & 19.8\\
\hline
\cellcolor{color13TeV} $A_{13}^{\phi Z}$& $gg\to H^{0}_{3}\to H^{0}_{1}Z \to (bb) Z$ & ATLAS & \cite{Aaboud:2018eoy} & [0.13;0.8] & 36.1\\
\hline
\cellcolor{color13TeV} $A_{13b}^{\phi Z}$& $bb\to H^{0}_{3}\to H^{0}_{1}Z \to (bb) Z$ & ATLAS & \cite{Aaboud:2018eoy} & [0.13;0.8] & 36.1\\
\hline
\end{tabular}
\caption{Neutral heavy Higgs boson searches at the LHC relevant for the GM scalars with final states including Higgs bosons. $\phi^0=H_1^0,H_3^0,H_5^0$, $\phi^{0\prime}=H_1^0,H_3^0$, $V=W,Z$ and $\ell=e,\mu$.}
\label{tab:directsearchesC}
\end{center}
\end{table}

\begin{table}
\begin{center}
\begin{tabular}{| l | l | l  l | c | c | c |}
\hline
\textbf{Label} &\textbf{Channel} & \multicolumn{2}{| l |}{\textbf{Experiment}} & \textbf{Mass range} & ${\cal L}$ \\
               &                 & \multicolumn{2}{| l |}{} &  \textbf{[TeV]} & \textbf{[fb$^{-1}$]} \\[1pt]
\hline
\hline
\cellcolor{color8TeV} $A_{8}^{\tau\nu}$ & $pp\to H_3^\pm \to \tau^\pm \nu $ & ATLAS &\cite{Aad:2014kga} & [0.18;1] & 19.5\\
\hline
\cellcolor{color8TeV} $C_{8}^{\tau\nu}$  & $pp\to H_3^+ \to \tau^+ \nu $ & CMS &\cite{Khachatryan:2015qxa}& [0.18;0.6] & 19.7\\
\hline
\cellcolor{color13TeV} $A_{13}^{\tau\nu}$  & \multirow{2}{*}{$pp\to H_3^{\pm} \to \tau^\pm \nu $ }& ATLAS & \cite{Aaboud:2018gjj} & [0.15;2] & 36.1\\
\cellcolor{color13TeV} $C_{13}^{\tau\nu}$ & & CMS & \cite{CMS-PAS-HIG-16-031} & [0.18;3] & 12.9\\
\hline
\hline
\cellcolor{color8TeV} $A_{8}^{tb}$ & $pp\to H_3^\pm \to t b $ & ATLAS & \cite{Aad:2015typ} & [0.2;0.6] & 20.3\\
\hline
\cellcolor{color8TeV} $C_{8}^{tb}$ & $pp\to H_3^+ \to t \bar{b} $ & CMS & \cite{Khachatryan:2015qxa} & [0.18;0.6] & 19.7\\
\hline
\cellcolor{color13TeV} $A_{13}^{tb}$ & $pp\to H_3^\pm \to tb $ & ATLAS & \cite{Aaboud:2018cwk} & [0.2;2] & 36.1\\
\hline
\hline
\cellcolor{color8TeV} $A_{8}^{WZ}$ & $WZ\to H_5^\pm \to WZ[\to(qq)(\ell\ell)]$ & ATLAS & \cite{Aad:2015nfa} & [0.2;1] & 20.3\\
\hline
\cellcolor{color13TeV} $A_{13}^{WZ}$ & \multirow{3}{*}{$WZ\to H_5^\pm \to WZ[\to(\ell\nu)(\ell\ell)]$} & ATLAS & \cite{Aaboud:2018ohp} & [0.2;0.9] & 36.1\\
\cellcolor{color13TeV} $C_{13,1}^{WZ}$ & & CMS & \cite{Sirunyan:2017sbn} & [0.2;0.3] & 15.2\\
\cellcolor{color13TeV} $C_{13,2}^{WZ}$ & & CMS & \cite{CMS-PAS-SMP-18-001} & [0.3;2] & 35.9\\
\hline
\hline
\cellcolor{color13TeV} $A_{13}^{4W}$ & $pp\to H_5^{\pm\pm}H_5^{\mp\mp}\to (W^\pm W^\pm)(W^\mp W^\mp)$ & ATLAS & \cite{Aaboud:2018qcu} & [0.2;0.7] & 36.1\\
\hline
\hline
\cellcolor{color8TeV} $C_{8}^{\ell^{\pm}\ell^{\pm}}$ & $VV\to H_5^{\pm\pm}\rightarrow W^{\pm}W^{\pm}[\to (\ell^{\pm}\nu) (\ell^{\pm}\nu)]$ & CMS & \cite{Khachatryan:2014sta} & [0.2;0.8] & 19.4\\
\hline
\cellcolor{color13TeV} $C_{13}^{\ell^{\pm}\ell^{\pm}}$ & $VV\to H_5^{\pm\pm}\rightarrow W^{\pm}W^{\pm}[\to (\ell^{\pm}\nu) (\ell^{\pm}\nu)]$ & CMS & \cite{Sirunyan:2017ret} & [0.2;1.0] & 35.9\\
\hline
\end{tabular}
\caption{Charged heavy Higgs boson searches at the LHC relevant for the singly and doubly charged scalars in the GM model. Again, $V=W,Z$ and $\ell=e,\mu$.}
\label{tab:directsearchesD}
\end{center}
\end{table}

We consider a large variety of direct searches for heavy resonances performed by the ATLAS 
and CMS Collaborations in Run 1 and Run 2 of the LHC. Tables \ref{tab:directsearchesA}, \ref{tab:directsearchesB}, \ref{tab:directsearchesC} and \ref{tab:directsearchesD} summarize the experimental searches to date which 
can have sensitivity to the neutral scalars $H^{0}_1$, $H^{0}_{3}$ and $H^{0}_{5}$ in 
the GM model. 
Tables \ref{tab:directsearchesA} and \ref{tab:directsearchesB} show all searches for a scalar resonance decaying into fermions or gauge bosons, and in Table~\ref{tab:directsearchesC} we list
the cases with decays including one or two Higgs bosons.
In Table~\ref{tab:directsearchesD}, we list all searches for singly and doubly
charged heavy scalars considered in our fits. Note that we are not sensitive in this model to 
the doubly charged Higgs searches in Refs.~\cite{Chatrchyan:2012ya,CMS-PAS-HIG-14-039,CMS-PAS-HIG-16-036}, where a $100\%$ branching fraction to leptons is assumed and the decay of $H^{\pm\pm}_5$ to $W^{\pm}W^{\pm}$ is suppressed, a scenario quite contrary to what we are considering here. The ATLAS searches for a doubly charged Higgs in Refs.~\cite{ATLAS:2014kca,Aaboud:2017qph} can have sensitivity in the two-lepton and three-lepton signal regions, and have been reinterpreted in the context of the Higgs triplet model \cite{Kanemura:2014ipa} and GM model \cite{Logan:2015xpa}. Also these limits are not applicable to our case due to the $\mathcal{B}(H^{\pm\pm}\to \ell^{\pm}\ell^{\pm})=100\%$ assumption, and, since in this work we are not formally recasting these searches, we choose not to include them in the fits.

The analyses in Tables \ref{tab:directsearchesA}, \ref{tab:directsearchesB}, \ref{tab:directsearchesC} and \ref{tab:directsearchesD} provide either model-independent $95\%$ confidence level upper limits on the production cross-section times branching ratios, $\sigma \cdot \mathcal{B}$, for different production and decay modes, or they are quoted by $\sigma \cdot \mathcal{B}/(\sigma \cdot \mathcal{B})_{\mbox{\tiny SM}}$ as a function of the resonance mass. If the experimental result includes the branching ratio into a specific final state in the upper limit, we write this channel using parentheses to combine particles which stem from a primary decay product. Whenever a secondary final state is given in square brackets, it means that we are quoting the limit on the primary final state measured through that particular secondary final state.

In order to assess which parts of the GM model parameter space 
are favored after imposing these constraints, we first calculate the theoretical 
production cross-section times branching ratio, $\sigma \cdot \mathcal{B}$, for all modes. 
For the neutral $H^{0}_1$, $H^{0}_3$ and singly charged $H^{\pm}_{3}$ states, we calculate $\sigma \cdot \mathcal{B}$ taking inputs from the two-Higgs doublet model already implemented in {\texttt{HEPfit}} \cite{Chowdhury:2017aav,Cacchio:2016qyh}, and rescale it to the GM model. We make use of the cross-section tables computed in Refs.~\cite{Chowdhury:2017aav,Cacchio:2016qyh} and calculate all branching ratios taking inputs from the couplings defined in the Appendix of Ref.~\cite{Hartling:2014zca}. For the VBF production cross-sections for $H^{\pm\pm}_{5}$, $H^{\pm}_{5}$, and $H^{0}_{5}$, we use the 8 TeV and 13 TeV production cross-section tables from the LHC Higgs Cross-Section Working Group~\cite{LHCXSWG}. The remaining VH quintet production modes and pair production of doubly charged $H^{\pm\pm}_{5}$ are calculated 
with \textsc{MadGraph5\_aMC@NLOv2.6.1}~\cite{Alwall:2014hca} at the leading order, taking the spectrum 
of the model generated with~\textsc{GMCalc}~\cite{Hartling:2014xma} as input. All the mentioned tables are interpolated linearly within \HEPfit.

In order to compare a specific $\sigma \cdot \mathcal{B}$ (calculated in each case as above) 
with the experimental upper limit, we define a ratio for the theoretical value and the observed 
limit, to which we assign a Gaussian likelihood with zero central value,
which is in agreement with the null results in the searches of heavy scalars so far.
The corresponding standard deviation of the Gaussian likelihood is adjusted in a way that the value of 1 for this ratio can be excluded with a probability of 95\%.

\section{Results}
\label{sec:results}

\begin{figure}
\centering
\includegraphics[width=\textwidth]{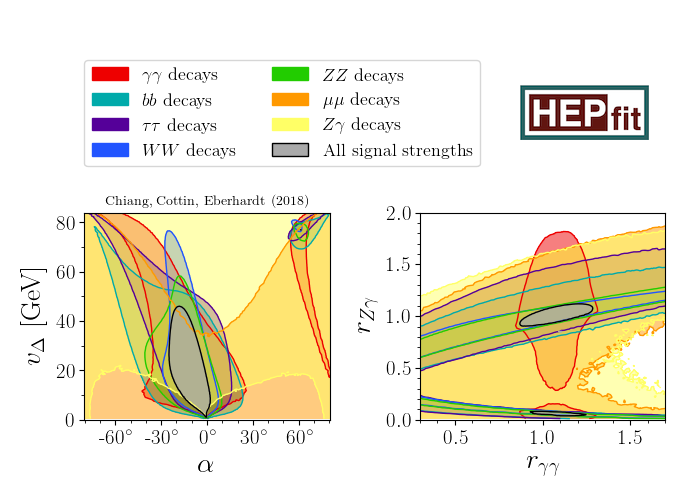}
\caption{Impacts of Higgs signal strengths on the $v_{\Delta}$-$\alpha$ plane (left) and on the relative one-loop couplings of $h$ to $\gamma\gamma$ and $Z\gamma$, $r_{\gamma\gamma}$ and $r_{Z\gamma}$, respectively (right). The $95\%$ probability contours are shown from fits to the data for $h$ decays to $\gamma\gamma$ (red), $Z\gamma$ (yellow) , $WW$ (blue), $ZZ$ (green), $bb$ (cyan), $\tau\tau$ (purple) and $\mu\mu$ (orange). The combined fit to all $h$ signal strengths is shown in grey.}
\label{fig:signalStrengths}
\end{figure}

Here we show the impact of all the constraints considered on the GM model. We first discuss the effect of the measured $h$ signal strengths. In Figure~\ref{fig:signalStrengths}, we show the individual impacts of specific decay categories on the $\alpha$-$v_{\Delta}$ plane and on the plane of the relative loop couplings of $h$ to $\gamma \gamma$ and $Z\gamma$, as well as the combination of all signal strengths. While the colored contours represent the allowed regions with $95\%$ probability for each decay mode, the grey region gives the combined fit.

\begin{figure}[t!]
\centering
\includegraphics[width=\textwidth]{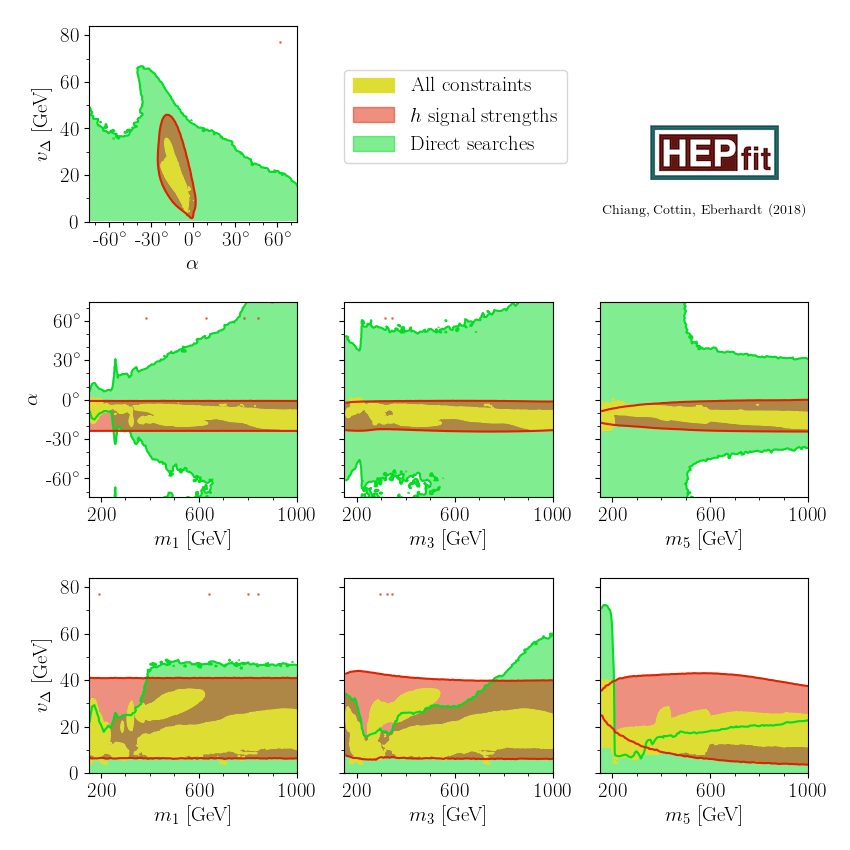}
\caption{Allowed $95\%$ probability regions in the $\alpha$ vs. $v_{\Delta}$, $\alpha$ vs. mass and $v_\Delta$ vs. mass planes. The red contour shows the effect of the $h$ signal strengths, while in green we show the impact of the direct searches. The combined fit with all constraints is shown in yellow.}
\label{fig:fitAll}
\end{figure}

Two allowed grey regions can be seen in the left panel of Figure \ref{fig:signalStrengths}. The bigger region close to $\alpha \approx 0^{\circ}$ (corresponding to the decoupling limit of the model) shows that $v_{\Delta}$ cannot exceed $\approx 45$ GeV, and negative $\alpha$ is mostly favored. The other allowed solution close to $\alpha\approx 61^{\circ}$ and $v_\Delta \approx 77$~GeV is only visible as a small black dot and features a negative sign for the $h$ couplings to vector bosons relative to the SM ($r_{ZZ}=r_{WW}=-1$). This region was not identified before as a viable possibility in the GM model (see, for instance, Ref.~\cite{Chiang:2015amq}), highlighting the advantages of using a global fitter. This region is visible only when considering the individual constraints of the $h$ signal strengths.  It disappears after taking into account the direct search results (see Figure \ref{fig:fitAll}). However, if one relaxes the assumptions about the considered GM mass ranges and the direct search constraints do not apply, this may persist as a viable scenario.
Compared to the parameter space in the $\alpha$-$v_{\Delta}$ plane given in Ref.~\cite{Chiang:2015amq}, the bigger allowed area here is smaller in size, as now we see that $\alpha$ cannot reach beyond $-25^{\circ}$. This is due to the much larger dataset on $h$ signal strengths made available in the recent years as well as the addition of the $\gamma \gamma$ signal strengths.

In the right panel of Figure \ref{fig:signalStrengths}, we show the $95\%$ probability contours in the $r_{Z\gamma}$-$r_{\gamma\gamma}$ plane, illustrating the impact on the one-loop couplings of $h$ to $\gamma\gamma$ and $Z\gamma$ relative to the SM. The information on the loop couplings is complementary to the tree-level couplings, which can be purely determined for a given pair of $\alpha$ and $v_\Delta$ from the left panel. We observe a solution around the SM values, while a much smaller $h$ coupling to $Z\gamma$ than the SM also remains allowed, since so far we only have upper limits on the $Z\gamma$ signal strength. Both are mainly determined by the $\gamma\gamma$, $WW$ and $ZZ$ final states.

\begin{figure}
\centering
\includegraphics[width=\textwidth,angle=0]{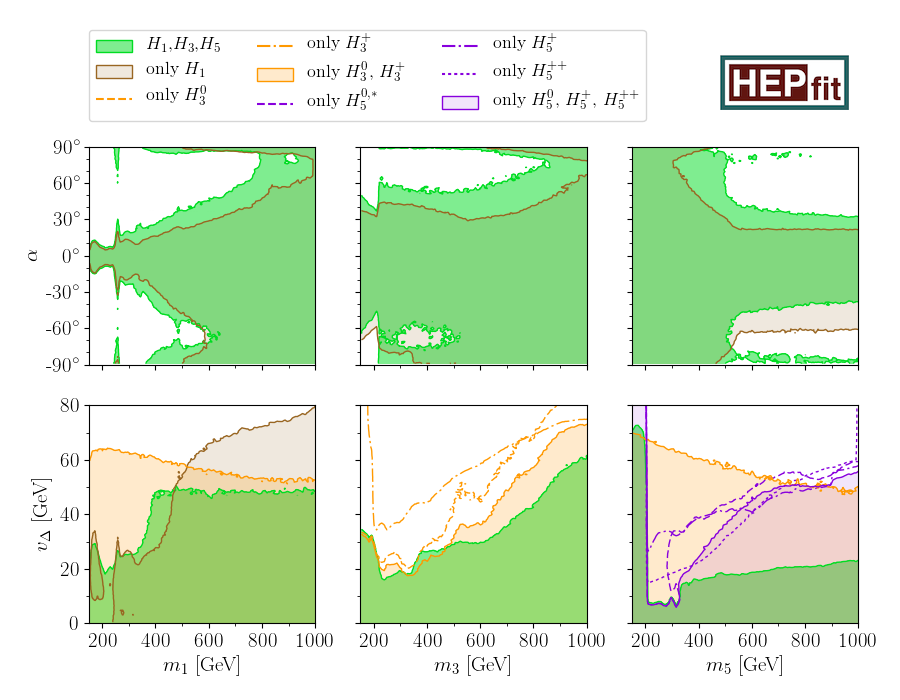}
\caption{Impact of different sets of direct searches on the $\alpha$ and $v_\Delta$ vs.~$m_{1,3,5}$ planes. The light brown/orange/purple shaded contours show the $95\%$ allowed regions after considering only searches for $H_1$/$H_3$/$H_5$ particles. For the searches for $H_3$ and $H_5$ resonances, we show the single contributions by the neutral and singly (doubly) charged scalar search limits with the dashed and dash-dotted (dotted) lines, where the region above the lines is excluded with a probability of $95\%$.}
\label{fig:directsearches}
\end{figure}

After the discussion of individual $h$ signal strengths, we want to have a glance at the the direct searches and their breakdown into searches for $H_1$, $H_3^0$, $H_3^+$, $H_5^0$, $H_5^+$ and $H_5^{++}$. In Figure \ref{fig:directsearches} we show their separate impacts on the mass-dependent $95\%$ limits on $\alpha$ and $v_\Delta$, as well as the combined fit to all of them. The angle $\alpha$ is only affected by the absence of $H_1$ signals, because the $H_3$ and $H_5$ couplings to fermions and gauge bosons do not depend on this parameter. The limits in the $\alpha$-$m_1$ plane are rather strong for relatively small masses, sometimes even stronger than the limits from the $h$ signal strengths, except for a small strip just above the kinematic $H_1\to hh$ threshold. Here this channel is not constrained by the $hh$ searches yet, and all the other branching ratios are sufficiently suppressed with respect to the $hh$ one to weaken the search constraints from the other decays.
With increasing mass, the $H_1$ search limits become less constraining, such that for instance for $m_1\gtrsim 600$ GeV all negative values of $\alpha$ are allowed. The difference between ``$H_1$ only'' and the ``all direct searches'' contours can be explained by a distortion of the allowed parameter space which is more obvious in the $v_\Delta$ vs.~mass planes. From the $m_1$ dependence of the triplet VEV limit one can see that where the $H_1$ searches become weak, the impact of searches for $SU(2)_V$ triplet bosons takes over. A detailed insight about the contribution of neutral and charged $H_3$ limits can be found in the middle panels; we learn that the former are more important if $m_3<800$ GeV. For heavier $H_3$ particles, all corresponding search limits are relatively weak. The same breakdown into the impact of neutral and charged $H_5$ resonance searches can be found in the right panels of Figure \ref{fig:directsearches}, here separately showing the roles of the singly and doubly charged search limits. While for $m_5<200$ GeV the experimental data on $H_5$ searches are not constraining at all, they yield the strongest restriction of all searches on $v_\Delta$ for quintet masses between 200 and 330 GeV, where the $H_5^0$ constraints are dominant. For the $m_5$ range from there up to 600 GeV, the three search analyses for doubly charged scalars are the most constraining.  Above these mass scales all searches are more or less equally important, even if the limits on the triplet VEV are much weaker than around 250 GeV.

We want to stress that up to this point the strongest upper limit on $v_\Delta$ for $m_5<200$~GeV stems from $H_3$ searches and it is only at around 70 GeV. There seems to be a lot of parameter space left to exclude by searches for low-mass $H_5$ resonances (between 125 and 200 GeV), which decay to one real and one virtual gauge boson. This is a measurement that should easily be performed by the LHC experiments, and it could significantly lower the upper limit on the triplet VEV in the global fit.

We should also mention that even though our mass priors go up to 1.1~TeV in the fits, we have decided to only show the regions up to 1~TeV.  This is because very close to the upper prior boundaries we observe that the contours become artificially small, and we do not want to confuse readers with this possibly misleading information.

Having scrutinized the $h$ signal strengths and the absence of direct search signals individually, let us move to their combination with the theoretical bounds.
In Figure~\ref{fig:fitAll}, we see the effects of individual sets as well as all constraints in the $v_{\Delta}$-$\alpha$ plane (top row), the $\alpha$-$m_{1,3,5}$ plane (middle row), and $v_{\Delta}$-$m_{1,3,5}$ plane (bottom row). After considering all constraints, the ``wrong sign'' region from Figure~\ref{fig:signalStrengths} gets excluded by the direct searches: In the middle row, we see that even if this exotic solution seems to be compatible with $H_1$ searches and not too far away from the regions allowed by $H_3$ searches, there are no red dots around $\alpha\approx 61^\circ$ in the $\alpha$-$m_5$ plane because all of the allowed points feature quintet masses above 1~TeV. This, however, is clearly in disagreement with the green contour stemming from all direct searches. In the combined fit including theory constraints, a rather constant region of $0^{\circ}\lesssim \alpha \lesssim 25^\circ$ is favored across the scanned mass ranges.
In the bottom row of Fig.~\ref{fig:fitAll}, we observe the interplay of the LHC observables and the bounds imposed by positivity and unitarity. Especially in the $v_\Delta$-$m_5$ plane, the contrast between the different sets of constraints becomes obvious, where a small region around $m_5\approx 250$ GeV and $v_\Delta\approx 13$ GeV seems to be excluded by both $h$ signal strengths and direct searches, but is ``resurrected'' in the global combination. Also, we see in the combination that the region in which $v_{\Delta}$ between 30 and $40$~GeV is allowed corresponds to $m_5<200$~GeV, where experimental improvements should be possible as mentioned above.

We note in passing that the decoupling limit~\cite{Hartling:2014zca} ($\alpha,v_{\Delta}\approx 0$) is not favored by the global fit due to the choice that our mass priors only go up to 1.1~TeV.

\begin{figure}[t!]
\centering
\includegraphics[width=400pt]{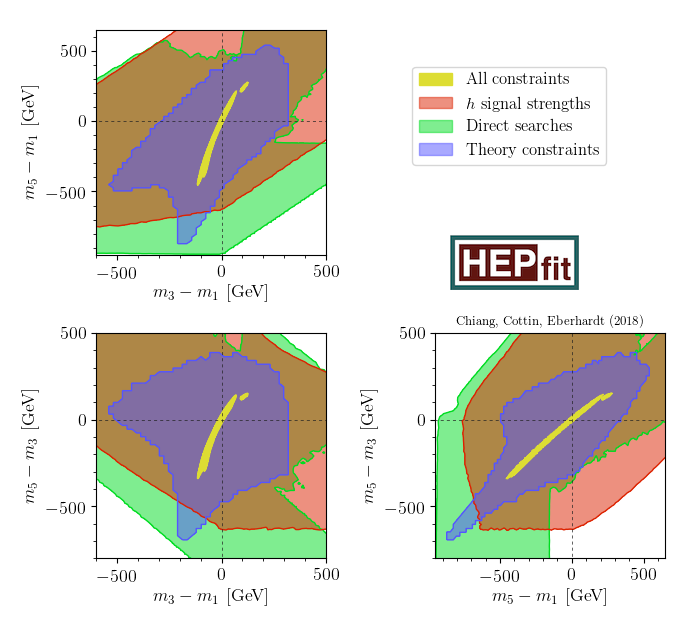}
\caption{Mass differences allowed at $95\%$ probability. We show regions in the planes of $m_{5}-m_{1}$ vs.~$m_{3}-m_{1}$ (top), $m_{5}-m_{3}$ vs.~$m_{3}-m_{1}$ (bottom left) and $m_{5}-m_{3}$ vs.~$m_{5}-m_{1}$ (bottom right). Effects of the theoretical constraints, Higgs signal strengths, and direct searches are shown in blue, red and green, respectively. The global fit with all constraints imposed is shown in yellow.}
\label{fig:fitAllmassdiff}
\end{figure}

\begin{figure}
\centering
\includegraphics[width=\textwidth]{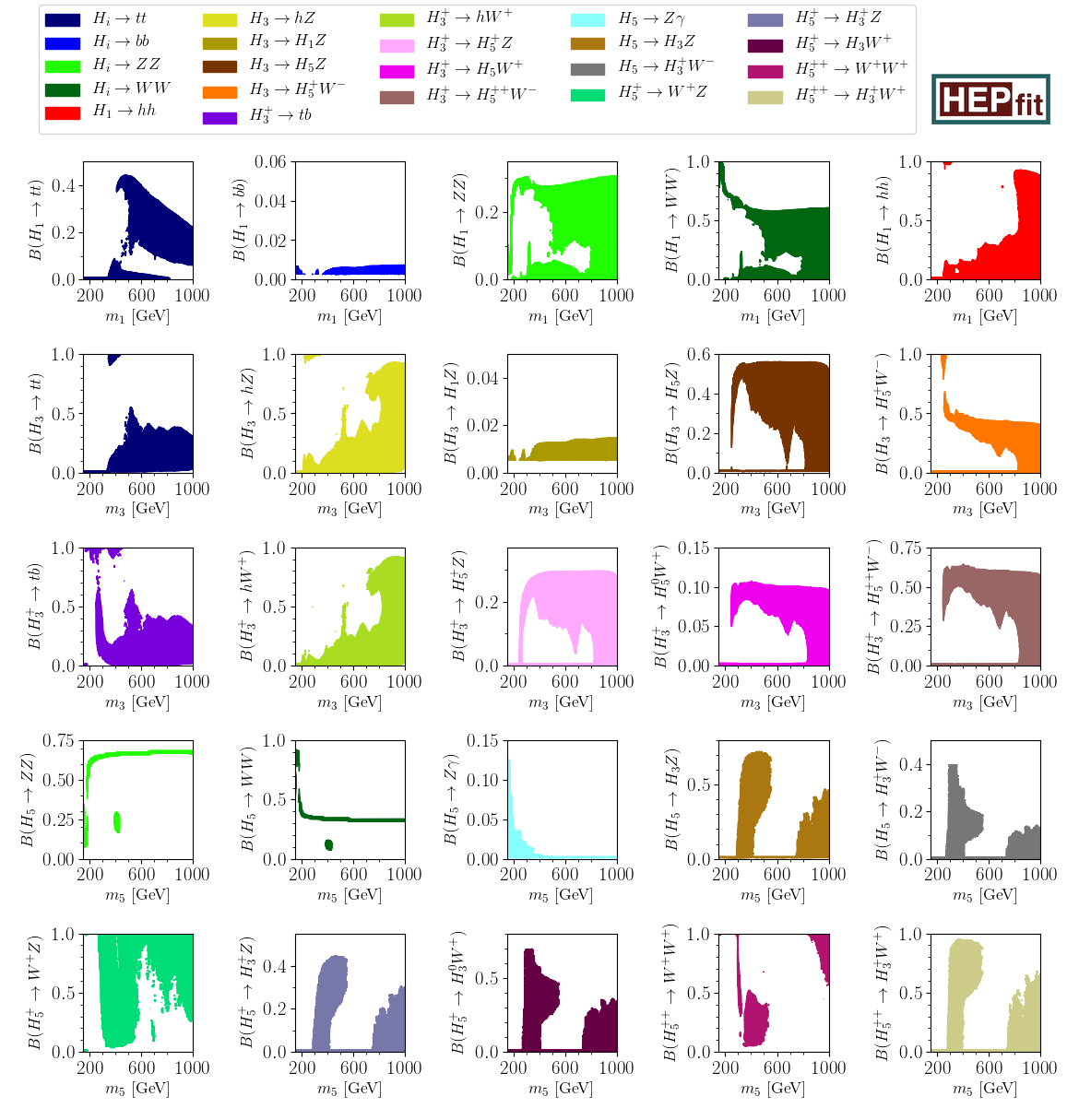}
\caption{95\% probability regions of the combined fit for the largest branching ratios of $H_1$ (top row), $H_3^0$ (second row), $H_3^+$ (third row), $H_5^0$ (fourth row) and charged $H_5$ (bottom row). Each color stands for a specific decay, and each of the neutral final states ($tt$, $bb$, $ZZ$ and $WW$) shares the same color among the $H_1$ and $H_3$ and $H_5$ bosons.}
\label{fig:fitBRs}
\end{figure}

Figure~\ref{fig:fitAllmassdiff} shows the effect of the theory bounds, $h$ signal strengths, direct searches and all constraints together on the mass differences of the exotic Higgs bosons in the model. Again, the colored contours represent the allowed regions with $95\%$ probability except for the theoretical constraints, for which we assume flat likelihoods. Hence, the $95\%$ contours would only reflect the prior shape, and the $100\%$ contours are used for theory.
Here we can see the power of the global fit: The individual sets of experimental constraints are not very strong in the $m_{5}-m_{1}$ vs.~$m_{3}-m_{1}$, $m_{5}-m_{3}$ vs.~$m_{3}-m_{1}$ and $m_{5}-m_{3}$ vs.~$m_{5}-m_{1}$ planes. The most dominant constraints come from the theoretical bounds, even though they still allow for a sizable region in the mass difference planes. However, once we combine the limits on $\alpha$ and $v_\Delta$ from the LHC experiments with the theoretical conditions in the global fit, the region that survives at 95\% shrinks to a thin strip for $|m_3- m_1|<150$ GeV (the yellow region). The disjoint regions at $m_5-m_1\approx 250$~GeV and $m_3-m_1\approx 120$~GeV are a consequence of our implementation of the direct searches: following Ref.~\cite{Hartling:2014xma}, we only include on-shell decays of the $H_5$ bosons. With an increasing $H_{5}$ mass, the decay to a neutral or charged $H_{3}$ and a massive vector boson can open abruptly. (The kinematic threshold is around $m_W$ plus the minimally allowed $m_3$, {\it i.e.}, at $\sim 230$~GeV.) For instance, the branching ratio of $H^{0}_{5}\to H^{0}_{3} Z$ can jump from zero to values over $50\%$ (see also Figure~\ref{fig:fitBRs}). If off-shell decays were also considered, the transition for these decays would become smoother and the two regions should be connected.

\begin{figure}
\centering
\includegraphics[width=\textwidth,angle=0]{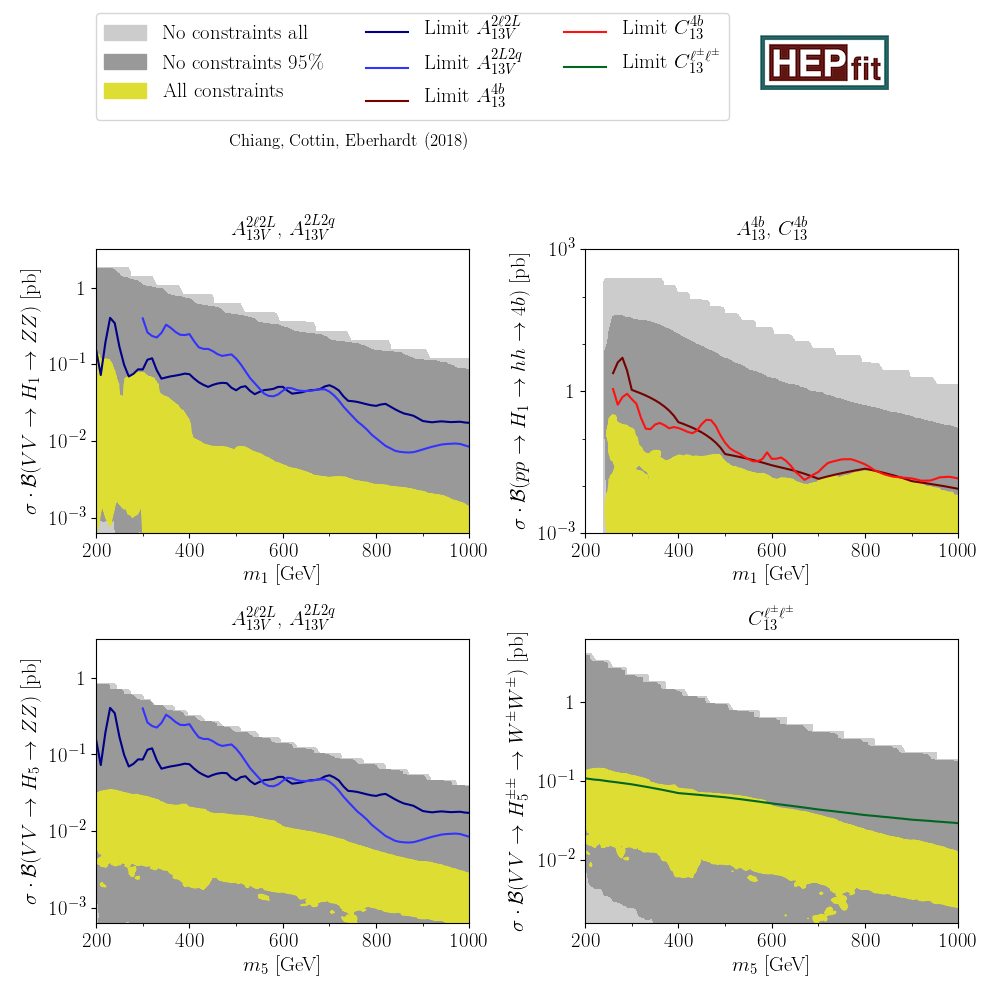}
\caption{Effect of five direct searches in the $\sigma \cdot \mathcal{B}$ vs.~mass planes. The light (dark) grey regions show the 100\% (95\%) prior regions. The allowed $95\%$ probability region after considering all the constraints is shown in yellow. The blue, red and green curves denote the experimental limits on $H_{1}$ or $H_{5}$ decays to two $Z$ bosons, on $H_1\to hh$ and on the pair production of $H_5^{\pm\pm}$.}
\label{fig:directS}
\end{figure}

After considering all the direct searches from the previous section, we find that the most powerful experimental analyses in constraining this model involve searches for the $H^{0}_{1,5}$ and $H^{\pm\pm}_{5}$ bosons. The effects of $H^{0}_{3}$ and $H^{\pm}_{3,5}$ observables are not as strong. In order to get more insights into our treatment of the direct searches and also to help the experimental collaborations better appreciate which search channels are more relevant or useful to the model, we show in Figure~\ref{fig:directS} five of the most constraining searches for heavy scalar resonances implemented in this work. They include the ATLAS and CMS searches for $VV\rightarrow H^{0}_{1,5}\rightarrow ZZ$ \cite{Aaboud:2017rel,Aaboud:2017itg} (labeled $A_{13V}^{2\ell2L}$ for fully leponically decaying $ZZ$ and $A_{13V}^{2L2q}$ for the semileptonic final state in Table \ref{tab:directsearchesB}), $pp\rightarrow H^{0}_{1}\rightarrow hh\rightarrow bbbb$~\cite{Aaboud:2018knk,Sirunyan:2018zkk} (labeled $A^{4b}_{13}$ and $C^{4b}_{13}$ in Table \ref{tab:directsearchesC}) and $VV\to H_5^{\pm\pm}\rightarrow W^{\pm}W^{\pm}$~\cite{Sirunyan:2017ret} (labeled $C_{13}^{\ell^{\pm}\ell^{\pm}}$ in Table \ref{tab:directsearchesD}).
The grey regions in the background delimit the available GM model space if we do not apply any constraint in the fit. We show the $100\%$ prior ranges, but also the $95\%$ prior regions, which differ only in the $H_1\to hh$ case by about one order of magnitude in $\sigma \cdot \mathcal{B}$.
All the five searches cut away a sizable portion of the allowed parameter space, ranging from a difference of less than one order of magnitude between the $H_5\to ZZ$ search limit and the grey contour to more than two orders of magnitude in $\sigma \cdot \mathcal{B}$ for the searches of $H_1\to hh$.
Comparing this to the role of the individual searches in the global fit (yellow contour), we observe that the searches in the left column are not very relevant except for $m_1<300$ GeV, while the channels in the right column yield the strongest constraint for $m_1$ between $500$ and $1000$~GeV and for $m_5$ between $200$ and $600$~GeV, respectively. (The reason why the 95\% allowed region in the last panel exceeds experimental exclusion limit for light $H_5$ can only be explained by the theoretical bounds that eliminate very low $\sigma \cdot \mathcal{B}$ values in the global fit. In the simultaneous fit to all direct searches only, the allowed contour stays below the CMS line.)

\begin{table}
\begin{center}
\begin{small}
\begin{tabular}[t]{| c | c || c | c |}
\hline
\textbf{Parameter} &\textbf{$95\%$ probability range}  & \textbf{Parameter} &\textbf{$95\%$ probability range }  \\[1pt]
\hline
\hline
$v_{\Delta}$  [GeV] \hspace{0.2cm}|\hspace{0.2cm} $\cos\beta$  &  $\leq 37$  \hspace{0.2cm}|\hspace{0.2cm} $\leq 0.42$ & $\lambda_1$ &  $[0.03 ; 0.22]$  \\
$\alpha$              &  $[-22^\circ ; -8^\circ]$    & $\lambda_2$ &  $[-0.65 ; 1.25]$    \\
$m_{5}-m_{3}$ [GeV]   &  $[-375 ; 125] $                 & $\lambda_3$ &  $[-0.9 ; 1.45]$    \\  
$m_{5}-m_{1}$ [GeV]   &  $[-500 ; 225]$                  & $\lambda_4$ &  $[-0.2; 0.65]$   \\  
$m_{3}-m_{1}$ [GeV]   &  $[-105  ; 105]$                  & $\lambda_5$ &  $[-3.0 ; 2.75]$    \\  
$g_{hhh}$ [GeV]   &  $[-455; 50]$                  & & \\  
\hline
\end{tabular}
\caption{ 95\% probability intervals of the GM model parameters after considering all the constraints in our fits, marginalizing over all other parameters.}
\label{tab:globalFitUL}
\end{small}
\end{center}
\end{table}

We present in Table~\ref{tab:globalFitUL} the $95\%$ probability ranges of the model parameters from our global fit. We do not get limits for the trilinear couplings $\mu_1$ and $\mu_2$. (More precisely, the limits that we observe in the fit are prior-dependent.) The upper limit of $105$~GeV on $m_1-m_3$ enables us to exclude the decays $H_1\to H_3^{0,+} H_3^{0,-}$, $H_1\to H_3 Z$ as well as $H_1\to H_3^+ W^-$ at the probability of $95\%$. In our fit, we also determine the 95\% allowed intervals for the triple $h$ coupling and the quartic couplings of the scalar potential. The SM value for the former is $g_{hhh}^{\text{SM}}\approx -190$ GeV.  In the GM model, it can be enhanced by a factor of 2.4 at most, but it can be also be very small and even have the opposite sign. The quartic couplings defined in \eqref{eq:GMpot} are mainly constrained by unitarity and positivity. While $\lambda_1$ and $\lambda_4$ cannot be larger than 0.65 in magnitude, $|\lambda_5|$ enjoys more freedom and can even be as large as 3.0 without violating the above-mentioned theory bounds.

\begin{table}
\footnotesize
\begin{center}
\begin{tabular}[t]{| c | c |}
\hline
\textbf{$H_{1}$}    & \textbf{$95\%$ prob.}   \\[0.5pt]
                    & \textbf{range}            \\[0.5pt]
\hline
\hline
$\Gamma_{1}$  &  $\leq$ 48 GeV                   \\
\hline
$\mathcal{B}(H^{0}_{1}\!\to\!tt)$           &  [0;45] $\%$ \\
$\mathcal{B}(H^{0}_{1}\!\to\!ZZ)$           &  [0;31] $\%$ \\
$\mathcal{B}(H^{0}_{1}\!\to\!WW)$           &  [0;100] $\%$ \\
$\mathcal{B}(H^{0}_{1}\!\to\!hh)$           &  [0;100] $\%$ \\
\hline
\end{tabular}
\begin{tabular}[t]{| c | c |}
\hline
\textbf{$H_{3}$} &\textbf{$95\%$ prob.}   \\[0.5pt]
                     &\textbf{range}    \\[0.5pt]
\hline
\hline
$\Gamma_{3}$  &  $\leq$ 70 GeV                   \\
\hline
$\mathcal{B}(H^{0}_{3}\!\to\!tt)$              & [0;100] $\%$ \\
$\mathcal{B}(H^{0}_{3}\!\to\!hZ )$             & [0;100] $\%$ \\
$\mathcal{B}(H^{0}_{3}\!\to\!H_{5}Z )$         & [0;56] $\%$ \\
$\mathcal{B}(H^{0}_{3}\!\to\!H^{+}_{5}W^{-})$  & [0;100] $\%$ \\
\hline
\hline
$\Gamma_{3+}$  &  $\leq$ 83 GeV                   \\
\hline
$\mathcal{B}(H^{+}_{3}\!\to\!tb)$               & [0;100] $\%$ \\
$\mathcal{B}(H^{+}_{3}\!\to\!hW^{+})$           & [0;93] $\%$ \\
$\mathcal{B}(H^{+}_{3}\!\to\!H^{+}_{5}Z)$       & [0;30] $\%$ \\
$\mathcal{B}(H^{+}_{3}\!\to\!H_{5}W^{+})$       & [0;11] $\%$ \\     
$\mathcal{B}(H^{+}_{3}\!\to\!H^{++}_{5}W^{-})$  & [0;64] $\%$ \\    
\hline
\end{tabular}
\begin{tabular}[t]{| c | c |}
\hline
\textbf{$H_{5}$} &\textbf{$95\%$ prob.}   \\[0.5pt]
                     &\textbf{range}    \\[0.5pt]
\hline
\hline
$\Gamma_{5}$  &  $\leq$ 18 GeV                   \\
\hline
$\mathcal{B}(H^{0}_{5}\!\to\!ZZ)$              & [7;68] $\%$ \\
$\mathcal{B}(H^{0}_{5}\!\to\!WW)$              & [7;93] $\%$ \\
$\mathcal{B}(H^{0}_{5}\!\to\!Z\gamma)$      & [0;13] $\%$ \\
$\mathcal{B}(H^{0}_{5}\!\to\!H_{3}Z)$          & [0;73] $\%$ \\
$\mathcal{B}(H^{0}_{5}\!\to\!H^{+}_{3}W^{-})$  & [0;40] $\%$ \\
\hline
\hline
$\Gamma_{5+}$  &  $\leq$ 18 GeV                   \\
\hline
$\mathcal{B}(H^{+}_{5}\!\to\!ZW^{+})$          &  [4;100] $\%$\\
$\mathcal{B}(H^{+}_{5}\!\to\!H^{+}_{3}Z)$      &  [0;46]  $\%$\\ 
$\mathcal{B}(H^{+}_{5}\!\to\!H^{0}_{3}W^{+})$  &  [0;70]  $\%$\\ 
\hline
\hline
$\Gamma_{5++}$  &  $\leq$ 18 GeV                   \\
\hline
$\mathcal{B}(H^{++}_{5}\!\to\!W^{+}W^{+})$        &  [5;100] $\%$\\
$\mathcal{B}(H^{++}_{5}\!\to\!H^{+}_{3}W^{+} )$   &  [0;95]  $\%$\\
\hline
\end{tabular}
\caption{ 95\% probability intervals of the GM model decay widths and branching ratios after considering all constraints in our fits. We only quote branching ratios larger than 5$\%$.}
\label{tab:globalFitUL2}
\end{center}
\end{table}

Limits on experimentally relevant derived quantities such as total decay widths and branching ratios for the $H_{1,3,5}$ scalars are presented in Table \ref{tab:globalFitUL2}. For the total decay widths of the heavy singlet, triplet and quintet particles, we observe that they cannot exceed $48$~GeV, $70$~GeV and $18$~GeV, respectively. For the SM-like Higgs boson, we obtain a probability range on $\Gamma_{h}$ between 3.9 and 4.5~MeV. In addition to the given branching ratio ranges of Table \ref{tab:globalFitUL2}, we also illustrate the mass dependence of the most important branching ratios in Figure \ref{fig:fitBRs}. Some of the planes contain white spaces in between allowed regions (in the ${\cal B}(H_1\to ZZ)$-$m_1$ plane, for instance, intermediate values are not completely filled at the 95\% level) because it is more likely that the decay channel is completely closed or open. The branching ratio values in between would require quite some fine tuning of the GM model parameters; nevertheless, they are not excluded. Here only the total upper and lower limits of the branching ratios and their mass dependence are important.
The discussed limits on the heavy Higgs decays can serve as a guidance for the LHC experiments in the design of new searches for the scalars in the GM model.

\section{Summary and outlook}
\label{sec:summary}

We have performed global fits in the Georgi-Machacek model for the first time, making  
use of the {\texttt{HEPfit}} package and the latest experimental data. We consider constraints from both theory (stability of the scalar potential and perturbative unitarity) and LHC Higgs observables. These include several up-to-date experimental results from Run 1 and Run 2 of the LHC, including all the data on Higgs boson signal strengths and eighty searches sensitive to the neutral, singly charged and doubly charged heavy Higgs particles of the Georgi-Machacek model. 

By considering only the signal strengths for the SM-like Higgs boson, we have found a previously unexplored region in the $v_{\Delta}$-$\alpha$ plane, featuring a negative sign in the Higgs couplings to vector bosons with respect to the SM couplings. This solution around $v_{\Delta}\approx 77$ GeV and $\alpha\approx 61^{\circ}$ cannot be ruled out by the signal strength data alone, but disappears as soon as direct search constraints are also imposed in the fit. However, with different assumptions on the masses of the heavy scalars this exotic solution might still be allowed.

The LHC searches for scalar resonances, especially the hunt for CP-even particles, constrain the vacuum expectation value of the Higgs bitriplet fields, but this upper limit could probably even be stronger if LHC data on $H_5$ searches below 200 GeV was available. Inclusive LHC diphoton searches could also help constrain the Georgi-Machacek model for low mass $H_{5}$, where Drell-Yan Higgs pair production
is sizable~\cite{Vega:2018ddp,Delgado:2016arn}.

Combining the LHC bounds with the theory constraints in a global fit, we extract 95\% probability limits on several Georgi-Machacek parameter regions and phenomenologically relevant quantities, which are significantly stronger than the bounds one would obtain when applying only one of the aforementioned sets of constraints. Among these are that $\alpha$ has to be between $-22^\circ$ and $-8^\circ$ and $v_\Delta$ smaller than $37$ GeV. The latter means that $\cos\beta$ cannot exceed $0.42$, which corresponds to an upper bound on $\sin \theta_H$, where the mixing angle $\theta_H$ is also used in the literature. We have found $95\%$ limits on the differences between the heavy Higgs masses of values less than $500$ GeV. The possibility of an $H_1$ decaying to $H_3$ can be excluded. We obtain upper $95\%$ bounds on the total decay widths of the Higgs states and on many branching ratios, for the latter even mass dependent limits. For instance, the $H_5^{\pm\pm}$ boson cannot decay into two $H_3^\pm$ bosons.

The existence of singly charged $H^{\pm}_{5}$ and doubly charged $H^{\pm\pm}_{5}$ scalars is a distinctive feature of the Georgi-Machacek model. Ongoing searches at the LHC (see Tables \ref{tab:directsearchesA} to \ref{tab:directsearchesD}) directly constrain $v_{\Delta}$. Current searches for $H^{\pm\pm}_{5}$ producing di-lepton resonances, which we did not consider in this work, could also be useful in constraining the Georgi-Machacek model in a global fit. This motivates flexibility in the definition of the experimental benchmarks in these searches to cases where the branching ratio of $H^{\pm\pm}_{5}$ to leptons is small, and its decay to vector bosons dominates. Other searches may also help to constrain the model when considering one-loop decays, such as the one proposed in~\cite{Logan:2018wtm} for $H^{\pm}\to W^{\pm}\gamma$.

This first global fit of the Georgi-Machacek model only represents the LHC part of the existing model constraints. Observables from flavor or electroweak precision physics could be used to further constrain the model. Also the destabilization of the custodial symmetry under renormalization group evolution is another interesting feature worthy of a detailed examination. In this context, we want to advertise the open-source \HEPfit package that can be used to address these and other questions in comprehensive statistical analyses.

\acknowledgments{The research of C.-W. C. was supported in part by the Ministry of Science and Technology (MOST) of Taiwan under Grant No. MOST-104-2628-M-002-014-MY4. G.C. acknowledges support by Grant No. MOST-106-2811-M-002-035. This work has been supported in part by the Agencia Estatal de Investigaci\'on (AEI, ES) and the European Regional Development Fund (ERDF, EU) [Grants No. FPA2014-53631-C2-1-P, FPA2017-84445-P and SEV-2014-0398]. We thank Jorge de Blas for help with the Higgs signal strength inputs and Ayan Paul for his technical support with {\texttt{HEPfit}}. We also thank Heather Logan for a discussion of our results and help with the {\texttt{GMcalc}} setup. The fits were performed on the cluster of INFN Roma Tre.}

\bibliographystyle{JHEP}
\bibliography{GMfit}

\end{document}